\documentclass[preprint,showpacs,preprintnumbers,amsmath,amssymb]{revtex4}

\usepackage{graphicx,amsfonts}
\usepackage{epsfig}
\usepackage{dcolumn}
\usepackage{bm}
\begin{document}

\title{Quasi-Dirac neutrinos in a model with local $B-L$ symmetry 
}

\author{A. C. B. Machado}%
\email{ana@ift.unesp.br}
\affiliation{ Instituto  de F\'\i sica Te\'orica--Universidade Estadual Paulista \\
R. Dr. Bento Teobaldo Ferraz 271, Barra Funda\\ S\~ao Paulo - SP, 01140-070,
Brazil }
\author{V. Pleitez}%
\email{vicente@ift.unesp.br}
\affiliation{ Instituto  de F\'\i sica Te\'orica--Universidade Estadual Paulista \\
R. Dr. Bento Teobaldo Ferraz 271, Barra Funda\\ S\~ao Paulo - SP, 01140-070,
Brazil }

\date{21/11/2012}
%
\begin{abstract}
In a model with $B-L$ gauge symmetry, right-handed neutrinos may have exotic local $B-L$ charge assignment:
two of them with $B-L=-4$ and the other one having $B-L=5$. Then, it is natural to accommodate the  right-handed neutrinos
with the same $B-L$ charge in a doublet of the discrete $S_3$ symmetry, and the third one in a singlet. If the
Yukawa interactions involving right-handed neutrinos are invariant under $S_3$, the quasi-Dirac neutrino
scheme arise naturally in this model. However, we will show how in this scheme it is possible to give a value for $\theta_{13}$ in
agreement with the Daya Bay results. For example the $S_3$ symmetry has to be broken in the Yukawa interactions involving right-handed
charged lepton.

 \end{abstract}


\maketitle

\section{Introduction }
\label{sec:intro}

Usually it is said that neutrino \emph{mass eigenstes} may be of the Dirac \emph{or} Majorana type. Notwithstanding, the
possibility that the definition of the particle and anti-particle is ambiguous was pointed out many years ago from two
different motivations. The first one, arises when Jauch~\cite{jauch} studying the quantization of spinor fields showed that
there are anticommutation relations which cannot follow from the Schwinger's action principle~\cite{schwinger}, and that it
implies the existence of fermion fields with an intermediate nature between those of the Dirac and of the Majorana fields.
Jauch showed that a  real parameter $\rho$ appears in the anticommutation relation between the spinor field at different
space-time points and that it may has values in the close interval $[0,1]$. The Dirac field corresponds to $\rho=0$,
while $\rho=1$ corresponds to the Majorana field. The cases with $0<\rho<1$ correspond to what is now called pseudo-Dirac
or quasi-Dirac fermion (see below). A Lagrangian formulation of this kind of fields was done in Ref.~\cite{japa}. Next, Pauli
\cite{pauli} and others \cite{enz}, considering the conservation of global charges as the lepton number, come to the same
conclusion about the existence of such a kind of fields. If interactions that violate the conservation of these charges are
allowed,  the neutrino field which enter in the weak interactions are of the form $a\nu+b\nu^c$, with $\vert a\vert^2+\vert
b \vert^2=1$. Moreover, Pauli pointed out the theoretical possibility that processes like the neutrinoless double beta decay
may have  cross sections with values between zero and the theoretical maximum value. The latter point was rediscovery in the
context of a gauge model in Ref.~\cite{valle}.

In modern gauge theories, since there are several neutrino flavors, these sort of fields can be realized in several ways.
They occur when two Majorana neutrinos are mass degenerated and have opposite parity so they are equivalent to one Dirac
neutrino~\cite{langacker}. However, the degenerescence would be remove by the weak interactions, when quantum corrections are
taken into account the would be a Dirac fermion will split into two Majorana neutrinos with different masses. But, if the mass
splitting is small this Dirac neutrino will became a \textrm{pseudo-Dirac} when one of the mass generated neutrinos is active and the
other is sterile~\cite{wolfenstein,doi} or \textrm{pseudo-Dirac} if both of them are active neutrinos~\cite{ valle,petcov,sarkar}.

More recently, a quasi-Dirac neutrino scheme was proposed in Ref.~\cite{moha10}. This model uses the $S_3$ discrete symmetry
to generate, at the tree level, the so called tribimaximal mixing matrix in the lepton sector. This was also implemented in a model
whith exotic right-handed neutrinos i.e., they carry non-convencional $U(1)_{B-L}$ charges~\cite{blsm}. The aim of the present paper
is to illustrate the main characteristics of the scheme and its limitations to produce a $\theta_{13}\not=0$ in agreement with the
recent Daya Bay results~\cite{dayabay}. Solar neutrino data may impose too strong constraints that it is not possible to generated
through quantum corrections the observed $\theta_{13}$.

The outline of this paper is as follows: In Sec.~\ref{sec:exotic} we review the model in which the scheme is implemented.
In Sec.~\ref{sec:cl} we consider the mechanism for generating charged lepton masses and in Sec.~\ref{sec:pmns} we show
how the observed $\theta_{13}$ could be generated. The last section is dedicated to our conclusions and some remarks.

\section{Models with non-identical right-handed neutrinos}
\label{sec:exotic}

The possibility that all neutrino \emph{flavor eigenstates} are part Dirac and part Majorana, as a consequence of a $S_3$ symmetry,
has been put forward recently~\cite{moha10}: two mass eigenstates are Majorana fields and one is a Dirac field at the tree level.
This scheme called bimodal/schizophrenic is just an interesting example of a quasi-Dirac neutrino based on
the $S_3$ symmetry. It is this symmetry that allow to distinguish among lepton generations (doublets and right-handed neutrinos).
However, if all right-handed neutrinos have the same quantum number their separation in $S_3$ irreducible representations is arbitrary.
This would not be the case if the  model has two right-handed neutrinos having a different quantum number from that of the
third one. It happens in the model of Ref.~\cite{blsm} because of the gauged $B-L$ symmetry. In fact, the
quasi-Dirac scheme was already implemented in a model~\cite{schinus} with local $B-L$ symmetry which has a different
scalar content with respect to that of Ref.~\cite{blsm}.

Here we will be concerned on how the non-zero $\theta_{13}$ may be obtained in the scheme which is also consistent with the Daya
Bay results~\cite{dayabay}. First, let us briefly review the main feature of the model which
has the following gauge symmetries:
\begin{eqnarray}
SU(3)_C\otimes SU(2)_L\otimes U(1)_{Y^\prime}\otimes U(1)_{B-L}
\nonumber \\ \downarrow  \langle\phi\rangle \nonumber \\
SU(3)_C\otimes SU(2)_L\otimes U(1)_Y \nonumber \\ \downarrow
\langle\Phi\rangle\nonumber
\\ SU(3)_C\otimes U(1)_{em},
\label{group}
\end{eqnarray}
where $Y^\prime$ is chosen in order to
obtain the hypercharge $Y$ of the standard model, given by
$Y=~Y^\prime+~(B-L)$. Here $\langle\phi\rangle$ and $ \langle\Phi\rangle$ denote one (or several) singlets and
doublets of $SU(2)_L$, respectively. Thus, in this case, the charge operator is given by
\begin{equation}
    \frac{Q}{e}=I_3+\frac{1}{2}\,\left[Y^\prime + (B-L)\right].
    \label{gn}
\end{equation}

The anomaly cancelation is also implemented if, for quarks, charged leptons and active
left-handed neutrinos the \textit{local} $B-L$ charges are as the usual ones, but for right-handed neutrinos
this charge is, instead of the usual assigment, $B-L=-4$ for two of them and $B-L=5$ for the third one. For this reason we call
them \textit{exotic} right-handed neutrinos. Thus, these neutrinos are, because of the local
$B-L$ and $Y^\prime[=-(B-L)]$, charges, naturally split in the two irreducible representation of $S_3$: $\textbf{3}
=\textbf{2}\oplus\textbf{1}$. After the breaking of the electroweak  gauge symmetry as is shown in Eq.~(\ref{gn}),
the usual \textit{global} $U(1)_B$ and $U(1)_L$ appear as  accidental symmetries as in the standard model. Also,
because of the non-standard local charges, the lepton sector has its own scalar sector and no large hierarchy in
the Yukawa couplings is necessary.
In this paper, as we said before, we will consider more details of the model of Ref.~\cite{schinus} and suggest
two possibilities of how a non-zero $\theta_{13}$ can be obtained.

The right-handed neutrinos with $B-L=-4$ are in a doublet of $S_3$, say $(n_{e R},n_{\tau R}$, may be consider heavy,
and the singlet say $n_{\mu R}$, is the light right-handed neutrino.
It is also assumed, as in the model of Refs.~\cite{moha10,schinus,moha11} (here we called it case $(a)$),
$D=(L_1,L_3)=(1/\sqrt6)(2L_e-L_\mu-L_\tau),(1/\sqrt2)(L_\mu-L_\tau))$ transforms as a
doublet and $L_S=L_2=(1/\sqrt3)(L_e+L_\mu+L_\tau)$ as a singlet. The scalar sector differs slightly from that of \cite{blsm}:
two scalar doublets with weak hypercharge $Y=-1$ are denoted by $\Phi_{1,2}=(\varphi^0_{1,2}\,\varphi^-_{1,2})^T$
are singlet of $S_3$. The mixing angles in the $(n_{eR},n_{\tau R})$ sector have been absorbed in $h_2$ and $h_3$. There
are also scalar singlets of $SU(2)$. For the quantum numbers of these fields see Ref.~\cite{schinus}. We denote
$\langle\varphi^0_1(\varphi^0_2)\rangle=v_1(v_2)$.

Therefore, after integrating the heavy degrees of freedom, the effective  Yukawa interactions that give neutrino masses are
\begin{equation}
-\mathcal{L}^{\textrm{eff}}_\nu=h_1\bar{L}_2\Phi_1n_{\mu R}\!+\!
\frac{h^2_2}{m_{n_e}}\,[\overline{(L^c_{1})_R}\,\Phi^*_2][L_{1L}\Phi^*_2]\!+\!
\frac{h^2_3}{m_{n_\tau}}\,[\overline{(L^c_{3})_R}\, \Phi^*_2][L_{3L} \Phi^*_2]
+H.c..
\label{yuka}
\end{equation}

From  the Yukawa interactions in Eq.~(\ref{yuka}), the neutrino mass matrix in the basis
\begin{equation}
\left(\begin{array}{cccc}
m_1&0&0&0\\
0&0&0& m_D\\
0&0&m_3&0\\
0&m_D&0&0
\end{array}\right)\;(a),
\label{casea}
\end{equation}
which has the eigenvalues $m_1,m_D,m_3,m_D$. At the tree level, there are four massive Majorana neutrinos, two of them,
an active and the sterile neutrino, are mass degenerated and correspond to a (quais)Dirac neutrino.

On the other hand, the neutrino mass matrix $\overline{(\chi^{\prime})} M^{0\nu}_M \chi^\prime$ written
in the active neutrino basis  $\chi^\prime=N^\prime_L+(N^\prime_L)^c$ where $N^\prime_L=(\nu_{eL},\nu_{\mu L},\nu_{\tau})^T_L$,
is of the form
\begin{equation}
M^{0\nu}_M=m_1\left(\begin{array}{ccc}
\frac{2}{3}&-\frac{1}{3}&-\frac{1}{3}\\
-\frac{1}{3}&\frac{1}{6}+\frac{m_3}{2m_1}&\frac{1}{6}-\frac{m_3}{2m_1}\\
-\frac{1}{3}&\frac{1}{6}-\frac{m_3}{2m_1}&\frac{1}{6}+\frac{m_3}{2m_1}
\end{array}\right),
\label{massa}
\end{equation}
The eigenvalues of the mass matrix (\ref{massa}) are $(m_1,0,m_3)$ and will be denoted by
$m^M_1=m_1$, $m^M_2=0$, and $m^M_3=m_3$. The massive Majorana neutrinos are $\nu_1$ and $\nu_3$, while $\nu_D$ has no
Majorana mass at tree level, and we have definded $m_1=h^2_2v^2_2/m_{n_e}$ and $m_3=h^2_3v^2_2/m_{n_\tau}$.
The matrix in Eq.~(\ref{massa}) is a consequence of the $S_3$ symmetry~\cite{moha06}. This matrix is diagonalized by the
tribimaximal matrix~\cite{harrison}:
\begin{equation}
V^\prime=U_{TB}\Omega^\prime =\left(\begin{array}{ccc}
\sqrt{\frac{2}{3}}& \frac{1}{\sqrt3}&0\\
-\frac{1}{\sqrt6}&\frac{1}{\sqrt3}&-\frac{1}{\sqrt2}\\
-\frac{1}{\sqrt6}&\frac{1}{\sqrt3}&\frac{1}{\sqrt2}
\end{array}\right) \Omega^\prime,
\label{tribi}
\end{equation}
where $\Omega^\prime=\textrm{diag}(e^{i\rho},1,1)$. The mass eigenstates basis denoted by
$N_L=(\nu_{1L},\,\nu_{DL},\,\nu_{3L})^T$ are related to the flavor basis as $N^\prime_L=U_{TB}N_L$. Notice that as a
result we have only one Majorana phase because one of the eigenvalues of the matrix (\ref{massa}) is zero.

We will  understand better this situation if we consider the $4\times4$ mass matrix, $\bar{\chi}M\chi$,
where $\chi_i=N_{iL}+(N_{iL})^c$, and $N_{iL}=(\nu_{1L}\,\nu_{2L}\,\nu_{3L}\,
n^c_{\mu L})^T$.  The eigenvalues of the matrix in Eq.~(\ref{casea}) are $m_1,m_D,m_3,m_D$ where $m_1,m_3$ are as given above, and
$m_D=h_1v_1/\sqrt2$. The respective eigenvectors are $\nu^M_1=\chi_1=(1,0,0,0)$, $\nu^M_2=(1/\sqrt2)(0,\chi_2,0,\chi_4)$,
$\nu^M_3=\chi_3=(0,0,1,0)$ and $\nu^M_4=(-i/\sqrt2)(0,\chi_2,0,-\chi_4)$, i.e., we have two Majorana neutrinos, $\nu^M_1,\nu^M_3$
and a Dirac neutrino formed by two mass degenerate Majorana neutrinos $\nu_D=\nu^M_2+i\nu^M_4$. The matrix in Eq.~(\ref{casea})
has a conserved lepton number induced by the transformation acting only on the fields $\chi_2\to e^{-i\beta}\chi_2$ and $\chi_4\to
e^{-i\beta}\chi_4$, implying the Dirac character of $\nu_D$, which is not conserved by the weak interactions, hence we have
a quasi-Dirac neutrino formed by an active and a sterile neutrino (according to the notation discussed in Sec.~\ref{sec:intro}).

Next, let us also consider, from the Yukawa interaction in Eq.~(\ref{yuka}), a $4\times4$ mass term for the neutrinos,
$\bar{\chi}^\prime M^{0\nu}\chi^\prime$, but now in the basis $\chi^\prime_i =N^\prime_{iL}+(N^\prime_{iL})^c$ where
$N^\prime_{iL}=(\nu_e\, \nu_\mu\, \nu_\tau\, n^c_\mu )^T_L$ and
$M^{0\nu}$ given by:
\begin{equation}
M^{0\nu}=m_1\left(\begin{array}{cccc}
\frac{2}{3}&-\frac{1}{3}\-\-&\- -\frac{1}{3} \- &\- \frac{m_D}{\sqrt{3}m_1} \\
-\frac{1}{3}&\frac{1}{6}+\frac{m_3}{2m_1}\-\-&\-\- \frac{1}{6}-\frac{m_3}{2m_1}\- &\- \frac{m_D}{\sqrt{3}m_1} \\
-\frac{1}{3}&\frac{1}{6}-\frac{m_3}{2m_1}\-\-&\-\- \frac{1}{6}+\frac{m_3}{2m_1}\- &\- \frac{m_D}{\sqrt{3}m_1} \\
\frac{m_D}{\sqrt{3}m_1}\-\-&\-\-  \frac{m_D}{\sqrt{3}m_1}\-\- &\-\- \frac{m_D}{\sqrt{3}m_1} & 0
\end{array}\right),
\label{massa2}
\end{equation}
where $m_1$ and $m_3$ are defined as in the matrix in (\ref{massa}) and $m_D$ as above. The eigenvalues of the matrix
(\ref{massa2}), denoted by $m^M_1e^{i\rho},m^M_2,m^M_3,m^M_4$, are
\begin{eqnarray}
m^M_1 = m_1,\;\;
m^M_2 = m^M_4 \equiv m_D,\;\;
m^M_3 = m_3,
\label{auto}
\end{eqnarray}
and we see that the two Majorana masses $m^M_1$ and $m^M_3$ are the same as in the case of the matrix in Eq.~(\ref{massa}).
The four Majorana massive neutrinos have been split as follows: two of them have different masses
and are purely Majorana fermions; the other two, which are mass degenerated, fuse to form a Dirac massive neutrino.
All of this is at tree level. After the breaking of the gauge $B-L$ symmetry nothing protects neutrinos to gain small
Majorana masses by  quantum corrections, in particular, the left- and right- components of the would be Dirac neutrino,
$\nu_D$. However, it may be rather small and this neutrino will continue to be, for all practical
purposes, a Dirac neutrino (see below).

The mass matrix in (\ref{massa2}) is diagonalized by the matrix
\begin{equation}
V\vert_{(a)} = \left(\begin{array}{cccc}
\sqrt{\frac{2}{3}} &  1/\sqrt{6} & 0 & -1/\sqrt{6} \\
- 1/\sqrt{6} &  1/\sqrt{6} & - 1/\sqrt{2} & -1/\sqrt{6} \\
-1/\sqrt{6} &   1/\sqrt{6} & 1/\sqrt{2} &  -1/\sqrt{6}\\
0 & 1/\sqrt{2} & 0 & 1/\sqrt{2}
\end{array}\right)\Omega,
\label{massa3}
\end{equation}
and $\Omega=\textrm{diag}(e^{i\rho},1,1,i)$, which can be rewritten as the following matrix
product~\cite{moha11}
\begin{equation}
V\vert_{(a)}=\left(
\begin{array}{cccc}
\sqrt{\frac{2}{3}}& \frac{1}{\sqrt3}&0 & 0\\
-\frac{1}{\sqrt6}&\frac{1}{\sqrt3}&-\frac{1}{\sqrt2} & 0 \\
-\frac{1}{\sqrt6}&\frac{1}{\sqrt3}&\frac{1}{\sqrt2} & 0 \\
0 & 0 & 0 & 1
\end{array}
\right)
\left(
\begin{array}{cccc}
1 & 0 &0 & 0\\
0 & \frac{1}{\sqrt2}&0 & -\frac{1}{\sqrt2} \\
0 & 0 &1 & 0 \\
0 & \frac{1}{\sqrt2} & 0 & \frac{1}{\sqrt2}
\end{array}
\right)\Omega.
\label{tribi2a}
\end{equation}

The relation of the mass eigenstates $\chi_i=N_{iL}+(N_{iL})^c$, with
$N_L=(\nu_1\,\nu_2\,\nu_3\,\nu_4)_L$ with the flavor eigenstates $\chi^\prime$ is
given by $\chi=V|_{(a)}\chi^\prime$, where $V|_{(a)}$ is given in Eq.~(\ref{massa3}) or (\ref{tribi2a}). The respective Majorana mass
eigenstates fields are $\nu^M_1=\nu_{1L}+(\nu_{1L})^c$ and $\nu^M_3=\nu_{3L}+(\nu_{3L})^c$ which have masses
$\tilde{m}_1=m_1e^{i\rho}$ and $m_3$, respectively, and the other two Majorana fields $\nu^M_2=\nu_{2L}+(\nu_{2L})^c$,
and $\nu^M_4=\nu_{4L}+(\nu_{4L})^c$ form a Dirac field, $\nu_D=\nu^M_2+i\nu^M_4$, with a Dirac mass term, $m_D$. We define
$\sqrt{2}\,\nu_{DL}=\nu_{2L} -i\nu_{4L}\equiv (n_\mu)^c_L$ and $\sqrt{2}\,(\nu^c_D)_R=(\nu_2)^c_R+i(\nu_4)^c_R\equiv
n_{\mu R}$. Note that the mass eigenstates $\nu_1$ and $\nu_3$, when written on the flavor basis,
do not have the contribution of the fourth neutrino $n_{\mu R}$. This is a prediction of the model at tree level, since
it is the rotation matrix in (\ref{tribi2a}) which determines this outcome.

With the masses in (\ref{auto}) we obtain
\begin{eqnarray}
&&\Delta m^2_{21}=m^2_D-m_1^2=\frac{h^2_1}{6}v^2_1-h^4_2\frac{v^4_2}{2m^2_{n_e}},\nonumber \\&&
\vert \Delta m^2_{32}\vert=\vert m_3^2-m^2_D\vert=  \left\vert\frac{h^4_3v^4_2}{2m^2_{n_\tau}}-\frac{h^2_1v^2_1}{6}
\right\vert
\approx \vert \Delta m ^2_{31}\vert=\vert m^2_3-m^2_1\vert.
\label{deltas}
\end{eqnarray}
Experimentally, $\Delta m^2_{21}=(7.59\pm0.20)\times10^{-5}\,\textrm{eV}^2$,
$\Delta m^2_{32}=(2.43\pm0.13)\times10^{-3}\,\textrm{eV}^2$~\cite{pdg}.
Just for an illustration, these values can be fitted by choosing, for the normal hierarchy
(here and below all parameters with dimension of mass are in eV),
$h_1=0.01,h_2=0.025,h_3=0.71$, $v_1=1.85,v_2=10^6$, and $m_{n_e}=m_{n_\tau}=10^{11}$,
the neutrino masses are $m_D=0.0107$, $m_1=0.0063$, and $m_3=0.0504$. We can also obtain the inverted mass hierarchy,
for instance by choosing $h_1=0.0045,h_2=0.07,h_3=0.025$, $v_1=1.9155,v_2=10^6$, and $m_{n_e}=m_{n_\tau}=10^{11}$,
the neutrino masses are  $m_D=0.0505$, $m_1=0.0498$, and $m_3=0.0062$. Notice that it is a prediction of the model the
existence of a light sterile neutrino, $n_{\mu R}$.

In fact, it is possible to choose different representations from that we have called case (a), which leads to the
effective interactions in Eq.~(\ref{yuka}), in such a way that the neutrino mass term written as ($\bar{N}MN$) becomes
\begin{equation}
\left(\begin{array}{cccc}
m_1&0&0& 0\\
0&m_2&0& 0\\
0&0&0&m_D\\
0&0&m_D&0
\end{array}\right)\;(b),\;\;\textrm{or}\;\;\quad \left(\begin{array}{cccc}
0&0&0& m_D\\
0&m_2&0& 0\\
0&0&m_3&0\\
m_D&0&0&0
\end{array}\right)\;(c).
\label{casebc}
\end{equation}

The model of \cite{moha10,schinus} corresponds to the case (a),  the cases (b) and (c) arise if we define
b): in the right-handed neutrinos sector $n_{\tau R}$ is now the singlet and $(n_{eR},n_{\mu R})$ the doublet,
and $L_S=L_3$ and $D=(L_1,L_2)$. c): $n_{e R}$ is the singlet and $(n_{\mu R},n_{\tau R})$ the doublet, and in
$L_S=L_1$ and $D=(L_2,L_3)$. A similar analysis to that doing for the matrix in Eq.~(\ref{casea}), follows for
matrices in Eq.~(\ref{casebc}).

Instead of (\ref{tribi2a}) we have for the cases (b) and (c), respectively:
\begin{equation}
V\vert_{(b)}=\left(
\begin{array}{cccc}
\sqrt{\frac{2}{3}}& 0&\frac{1}{\sqrt3} & 0\\
-\frac{1}{\sqrt6}&-\frac{1}{\sqrt2}&\frac{1}{\sqrt3} &0\\
-\frac{1}{\sqrt6}&\frac{1}{\sqrt2}&\frac{1}{\sqrt3} & 0 \\
0 & 0 & 0 & 1
\end{array}
\right)
\left(
\begin{array}{cccc}
1 & 0 &0 & 0\\
0 & 1 &0 &0  \\
0 & 0 &\frac{1}{\sqrt2} & -\frac{1}{\sqrt2}\\
0 & 0&\frac{1}{\sqrt2}  & \frac{1}{\sqrt2}
\end{array}
\right)\Omega,
\label{tribi2b}
\end{equation}
with $\Omega$ as in Eq.~(\ref{tribi2a}), and
\begin{equation}
V\vert_{(c)}=\left(
\begin{array}{cccc}
\frac{1}{\sqrt3}& \sqrt{\frac{2}{3}}&0 & 0\\
\frac{1}{\sqrt3}&-\frac{1}{\sqrt6}&-\frac{1}{\sqrt2} & 0 \\
\frac{1}{\sqrt3}&\frac{1}{\sqrt6}&\frac{1}{\sqrt2} & 0 \\
0 & 0 & 0 & 1
\end{array}
\right)
\left(
\begin{array}{cccc}
\frac{1}{\sqrt2} & 0 &0 & -\frac{1}{\sqrt2}\\
0 & 1&0 &0  \\
0 & 0 &1 & 0\\
 \frac{1}{\sqrt2}& 0  & 0 & \frac{1}{\sqrt2}
\end{array}
\right)\Omega^{\prime\prime},
\label{tribi2c}
\end{equation}
where $\Omega^{\prime\prime}=(1,e^{i\rho},1,i)$.

As we said before, the tribimaximal matrix diagonalized the neutrino mass matrix at leading order. Corrections
to that matrix may arise from quantum loop corrections~\cite{araki} (or evolution with the renormalization
group equations~\cite{rge}) and/or the mixing in the charge lepton sector. In fact, global neutrino data analysis
had already suggested that $\theta_{13}\not=0$~\cite{theta13}, and a first evidence that this is the case has
been obtained from the observation, at the 2.5$\sigma$ level, of the appearance
$\nu_\mu\to\nu_e$~\cite{t2k,morisi}. More recently, the Daya Bay results are more conclusive, at the 5.2$\sigma$
level they found $\sin^2 2\theta_{13} = 0.092 ± 0.016(stat) ± 0.005(syst)$~\cite{dayabay}, which implies
$0.13\leq \sin\theta_{13}\leq0.17$~\cite{global}. Finite quantum corrections to the mass matrix in
Eq.~(\ref{massa}) may be consider but they are strongly suppressed by solar neutrino data because of the active
to sterile neutrino oscillation present in this model~\cite{cirelli,gouvea}. It is for this reason that the would be
Dirac neutrino is for practical proposes a Dirac fermion. This implies a $V_{PMNS}$ matrix of the tribimaxiaml type.
The only way to obtain a realistic form of this matrix is to have a nondiagonal mass matrix in the charged lepton
sector and using $V_{PMNS}=U^{\dagger}_lU^\nu_L$. Here, $U^\nu_L$ means the $3\times3$ submatrices in Eqs.~(\ref{tribi2a}),
(\ref{tribi2b}) and (\ref{tribi2c}). Next, we have to consider how it is possible, in the context of
the present model, to obtain a realistic $V_{PMNS}$ matrix. Before doing it, let us consider the mass matrix of
the charged lepton sector.

\section{The charged lepton sector }
\label{sec:cl}

Let us suppose that the $S_3$ symmetry also constrains the Yukawa interactions in the charged leptons.
In this case, to obtain the charged lepton mass matrix we add three extra scalar doublets with $Y=+1$, denoted $\Phi_e,\Phi_\mu,\Phi_\tau$,
and using the complex representation for $S_3$ determined by the matrix \cite{ishimori}
\begin{equation}
U_\omega=\frac{1}{\sqrt3}\left(\begin{array}{ccc}
1 & 1 &1\\
1 & \omega & \omega^2\\
1&\omega^2 &\omega
\end{array}\right),
\label{magic}
\end{equation}
which change from the irreducible basis $\mathbf{1}$ and $\mathbf{2}$ to the reducible one $\mathbf{3}$ without changing the
product rules. In this basis $\mathbf{3}_L=(L_e,\,L_\mu,\,L_\tau)$,
$\mathbf{3}_R=(e_R,\, \mu_R,\,\tau_R)$ and $\mathbf{3}_S=(\Phi_e,\,\Phi_\mu,\,\Phi_\tau)$.
The charged lepton sector the Yukawa interactions can be written using
the four $S_3$ singlets formed by the direct products of three triplets
$(x_i,y_i,z_i)$, $i=1,2,3$: $x_1y_1z_1+x_2y_2z_2+x_3y_3z_3$,
$x_1y_2z_3+x_1y_3z_2$, $x_2y_1z_3+x_3y_1z_2$, and $x_2y_3z_1+x_3y_2z_1$:
\begin{eqnarray}
-\mathcal{L}_{l}&=&G[\bar{L}_e\Phi_e e_R+ \bar{L}_\mu\Phi_\mu \mu_R+\bar{L}_\tau \Phi_\tau \tau_R]+
H\bar{L}_e(\mu_R\Phi_\tau+\tau_R\Phi_\mu)\nonumber \\ &+&
F(\bar{L}_\mu \Phi_\tau+\bar{L}_\tau\Phi_\mu)e_R+I(\bar{L}_\mu\tau_R+\bar{L}_\tau \mu_R)\Phi_e+H.c.
\label{yuka2}
\end{eqnarray}
From (\ref{yuka2}) we obtain the most general mass matrix for charged leptons (here $\langle\varphi^0_l\rangle=v_l/\sqrt2$):
\begin{equation}
M_l=\frac{1}{\sqrt2}\left(\begin{array}{ccc}
Gv_e & H v_\tau & H v_\mu\\
Fv_\tau &Gv_\mu &Iv_e\\
F v_\mu& Iv_e& Gv_\tau
\end{array}\right).
\label{clmass}
\end{equation}

Using the following values for the parameters (dimensional parameters in MeV): $v_e=9.97$, $v_\mu=2038.56$, $v_\tau=34924.8$,
$G=0.05$ and $H=1.95\times10^{-4}$, $F=0.35\times10^{-5}$, $I=1.92\times10^{-2}$, we obtain $m_e=0.4894$, $m_\mu=102.155$ and
$m_\tau=1746.24$. The mass matrices  are a prediction of the model, hence they are valid at the energies at which all symmetries
of the model are realized, \textit{i.e.}, at the electroweak scale. For this reason, we use the running lepton masses at $\mu=M_Z$
which values were taken from Ref.~\cite{massas}. The $U^l_L$ unitary matrix diagonalize the Hermitian $M_lM^\dagger_l$ matrix, where
$M_l$ is the mass matrix in (\ref{clmass}). We obtain for the values of the parameter above:
\begin{equation}
U^l_L=\left(\begin{array}{ccc}
0.9976 & -0.0698 &-0.0002\\
-0.0698 & -0.9976 & 0.0007\\
0.0002& 0.0007 & 0.9999
\end{array}\right).
\label{magic2}
\end{equation}

We have consider several set of values for the parameters in the Yukawa interactions (\ref{clmass}) and all of them
give similar values for the entries of the matrix $U^l_L$ as in (\ref{magic2}).

\section{The PMNS mixing matrix}
\label{sec:pmns}
Notwithstanding, the full PMNS matrix also includes rotations in the charged
lepton sector, i.e., it is defined as $V_{PMNS}=U^{l\dagger}_LU_{TB}$. Here, $U_{TB}$ denotes any of the
$3\times3$ submatrices in Eqs.~(\ref{tribi2a}), (\ref{tribi2b}) and (\ref{tribi2c}).

Here, we compute the leptonic mixing matrix, $V_{PMNS}$, for the three cases in Eqs.~(\ref{tribi2a}), (\ref{tribi2b})
and (\ref{tribi2c}), obtaining,
\begin{equation}
\vert V_{PMNS}\vert_{(a)}\approx \left(\begin{array}{ccc}
0.7875& 0.6144 &0.0473\\
0.4617&0.5375&0.7056\\
0.4081&0.5775&0.7070
\end{array}
\right),
\label{tribi4a}
\end{equation}

\begin{equation}
\vert V_{PMNS}\vert_{(b)}\approx \left(\begin{array}{ccc}
0.5358& 0.8429 &0.0495\\
0.6158&0.3500 &0.7059\\
0.5777 &0.4087 &0.7066
\end{array}
\right),
\label{tribi4b}
\end{equation}
and
\begin{equation}
\vert V_{PMNS}\vert_{(c)}\approx \left(\begin{array}{ccc}
0.8429 & 0.0492  &0.5358 \\
0.3500&0.7049 &0.6158 \\
0.4087&0.7076&0.5777
\end{array}
\right),
\label{tribi4c}
\end{equation}
respectively. From data we have~\cite{global}:
\begin{equation}
\vert V_{PMNS}\vert_{exp}=\left(\begin{array}{ccc}
0.78-0.845 & 0.52-0.61 &0.13-017\\
0.40-0.58 & 0.39-0.65 & 0.57-0.80\\
0.19-0.43& 0.53-0.74 & 0.59-0.81
\end{array}\right).
\label{magic3}
\end{equation}
Comparing the three $V_{PMNS}$ in Eqs.~(\ref{tribi4a})-(\ref{tribi4c}) with the matrix in Eq.~(\ref{magic3}),
we see that there is no agreement with the measured $V_{PMNS}$.

There are a few ways to turn around this trouble. On the one hand, since the neutrino and charged lepton masses are at
the $Z$-peak~\cite{massas}, it means that the running of these values to the low energies have to be done still~\cite{rge}
in the context of the present model. On the other hand, radiactive corrections will induce a non-zero $\theta_{13}$.
Notwithstanding, it implies that the corrections to the neutrino mass matrix are not to small, but since in this case
there are active to sterile neutrino oscillations the mass corrections are strongly constrained by solar data~\cite{gouvea}
in such a way that no realistic value for $\theta_{13}$ arises.

Another possible way to overcome this difficulty  is by considering the $S_3$ symmetry to be break in the charged lepton
sector. Thus, a $S_3$ non-invariant Yukawa interactions given mass to the charged leptons is
\begin{equation}
-\mathcal{L}^l=\bar{L}_{iL}Y^l_{ij}l_{jR} \frac{v_{SM}}{\sqrt2}+H.c.
\label{masscl2}
\end{equation}
where $Y^l$ is an arbitrary $3\times 3$ matrix and $H$ denotes the  Higgs boson of the standard model which gives mass to
the quarks too. If we solve, simultaneously the following equations
\begin{equation}
V_{PMNS}=U^{l\dagger}_LU_{TB},\quad \frac{v^2_{SM}}{2}U^l_LY^l(Y^l)^\dagger (U^l_L)^\dagger=\textrm{diag}(m^2_e,m^2_\mu,m^2_\tau),
\label{sol}
\end{equation}
we obtain that: in case $(a)$ has solutions with complex Yukawa parameters, case $(b)$ has no solution and, case $(c)$ has
solution with real Yukawa parameters. Here we show only the latter case: with
\begin{equation}
U^l_L\vert_{(c)} =\left(\begin{array}{ccc}
0.9114 & 0.8346  &0.6669 \\
0.1356&-0.4644 &-0.5304 \\
0.3240&0.5457&0.4440
\end{array}
\right).
\label{tribi4d}
\end{equation}
which implies, from (\ref{sol}), that
\begin{equation}
Y^l\vert_{(c)} =\left(\begin{array}{ccc}
8.98 & 1.84  &-21.01  \\
27.886&18.51  &33.97  \\
-26.68&-37.22 &-27.14
\end{array}
\right).
\label{yij}
\end{equation}
and we obtain
\begin{equation}
\vert V_{PMNS}\vert_{(c)}=\left(\begin{array}{ccc}
0.79 & 0.56 &0.13 \\
0.53 & 0.65 & 0.71 \\
0.33& 0.58 & 0.70
\end{array}\right),
\label{pmnsc}
\end{equation}
which in agreement with (\ref{magic3}).

\section{Conclusions}
\label{sec:con}

The quasi-Dirac neutrino scheme of Refs.~\cite{moha10,schinus}  is interesting in its own. Although there is a light
sterile neutrino, its spectrum is not of the form "3+1", since the fourth massive neutrino is quasi degenerated with
one of the active neutrinos. Hence, the effect of the extra mass square difference might appear only in neutrinos
coming from long distance sources like  supernovas and not as a solution to some possible neutrino anomalies~\cite{white}.

The mass matrix in equation (\ref{clmass}) is the best mass matrix that can be obtained without violating the $S_3$
symmetry and contribute to the value of $\theta_{13} \not=0$. However, without violating the $S_3$ symmetry in the
charged lepton Yukawa interactions we cannot explain the Daya Bay data. since the right-handed charged lepton do not have
exotic values for the gauge $B-L$ symmetry, the $S_3$ symmetry is not naturally incorporated
in that interactions. Hence, the validity of $S_3$ symmetry only in the neutrino sector can be a prediction of the model
that made the $B-L$ symmetry a gauge symmetry and exotic right-handed neutrinos. On the other hand, if the charged lepton
interactions violate the symmetry  $S_3$, we can get a general mass matrix, as that in Eq.~(\ref{masscl2}) and, as we
have shown, it is possible to fit a  realistic PMNS mixing matrix in agreement with the Daya Bay's results.

Some final remarks are in order. 1) If a Majorana mass for the sterile neutrino, say $n_{\mu R}$, is allowed the
$44$ entry in the mass matrix (\ref{massa2}) is non-zero. Notwithstanding, the Majorana mass for the $n_{\mu R}$ arises
only from non-renormalizable interactions since the operator $\overline{(n_{\mu R})^c}n_{\mu R}$ is not allowed by
the $B-L$ attribution of the model and the $Z_3$ symmetries, see Ref. \cite{schinus} for the respective quantum numbers.
Hence, the operator with the lower dimension generating this mass term is one of dimension seven:
$\lambda_M(\Phi_{_{SM}}\epsilon\Phi_2\phi_{1(2)}\phi_x^*/ \Lambda^3)\overline{(n_{\mu R})^c}n_{\mu R}$. In this case
$M=\lambda_Mv_{_{SM}}v_2u_1v_x/\Lambda^3$. Just for illustration, if $\lambda_M\sim O(1)$,  $u_1\stackrel{<}{\sim} \Lambda$,
$M\approx v_{_{SM}}v_2v_x/\Lambda^2$ is rather small using $v_{_{SM}}\sim100$ GeV, $v_2=10^{-3}$ GeV and
$\Lambda=1$ TeV we have $M\approx 10^{-13}(v_x/\textrm{GeV})$. Moreover, $v_x$ is not necessarily  large  since
it is not the responsible for breaking the $B-L$ gauge symmetry.  We see that the $M$ value is rather small
and satisfies the constraint from solar data~\cite{gouvea}.

2) The present model involves Higgs scalar doublets which couple mainly to leptons and have small VEVs.  Moreover, the model has
also scalar singlets and some of them may have VEVs lower than the TeV scale. Tiny VEVs have been proposed before
\cite{ma} and after \cite{davison} the models in Ref.~\cite{schinus,blsm}. Then, the question concerning on the stability
of the tree level VEVs arises. This has been done recently~\cite{vevs} in the context of the particular 2DHM of
Ref.~\cite{davison}. This question is rather model dependent and a similar analysis in the context of the present model will
be considered elsewhere.

3) The model in Ref.~\cite{blsm} was proposed just as a new solution to the anomaly cancellation when $B-L$ is a local symmetry, and
it has interesting features by its own. For instance, beside implementing the bimodal scheme without fine tuning in the neutrino Yukawa
interactions, it is a model which also implement naturally the features of the so called leptophilic two Higgs doublet model
(L2DHM)~\cite{l2dhm} and the neutrino specific 2HDM one~\cite{davison,logan}. In the latter models
those Higgs doublets were introduced \textit{ad hoc}. In general the supersymmetric versions of this sort of models has interesting
features in accelerator physics~\cite{sher} and in cosmology~\cite{cosmo}. Recently, a model with quasi-Dirac and Majorana
neutrinos in the context of supersymmetric standard model with the extra symmetries  $S_4\otimes({Z_3})^3$ has been
proposed~\cite{morisi2}. However, we would like to stress that all these features, for instance doublets given masses just for neutrinos,
arise naturally when we consider the anomaly free $B-L$ gauge symmetry and they are not assumed \textit{ad hoc}.
This avoids extreme fine tuning in the Yukawa coupling of the Dirac neutrino as $h_1=0.01(0.0045)$ in the normal(inverted)
hierarchy instead $h_1\sim10^{-12}$ as in \cite{moha10,cosmo}.

4) Finally, we stress that the existence of such a very light sterile neutrino is a prediction of the model, and it is not
motivated by possible  anomalies observed in neutrino experiments~\cite{white}. Since it is (almost) mass degenerate with one of the active
neutrinos its  effect may be only observed in extragalactic neutrinos~\cite{moha11}.

\acknowledgments{One of the author (ACBM) was supported by CAPES and (VP) was partially
supported by CNPq and FAPESP.}

\appendix

\section{charged leptons}
\label{sec:l2}

The charged lepton masses in Ref.~\cite{moha10} is generated by a dimension five operator and it is diagonal at tree level. Thus the $V_{PMNS}$
is just the matrix that diagonalized the neutrino mass matrix and this is just the tribimaximal one.
To obtain those authors introduce three gauge singlet scalars $\sigma_e,\sigma_\mu$ and $\sigma_\tau$ these fields and the right-handed charged letons transform
like the left-handed doublets and for avoiding a general mixing it is necessary to impos $Z_n$ symmetries: $Z_{n,e}\otimes Z_{n,\mu}\otimes Z_{n,\tau}$
in such a way that the right-handed leptons transfom as $\omega^p_{e,\mu,\tau}$ and the singlets scalars as $\omega^{-p}_{e,\mu,\tau}$. This case has been rouled out by
recent neutrino data.

Case 1. The right-handed components trasfoorm as $e_S=e_R$, $e_D=(\mu_R,\tau_R)$ and the Yukawa coupling is given by
\begin{equation}
-\mathcal{L}_l=h_e\bar{L}_2e_SH+h_{\mu\tau}[\bar{L}_De_D]_1H+H.c.
\label{ea1}
\end{equation}
where $H$ is the usual SM Higgs doublet. This case is not favored because the $\mu$ and $\tau$ are mass degenerated and the
matrix which diagonalize the mass matrix is the tribimaximal, hence $V_{PMNS}=U^2_{TB}=\textbf{1}$. \\
Case 2. Now
\begin{equation}
e_S=\frac{1}{\sqrt3}(e_R+\mu_R+\tau_R), \;\;e_D=\left[\frac{1}{\sqrt6}(2e_R-\mu_R-\tau_R),\frac{1}{\sqrt2}(\mu_R-\tau_R)\right],
\label{ea2}
\end{equation}
with the Yukawa interactions as in Eq.~(\ref{ea1}). In this case the matrix which diagonalize the charged leptons mass matrix is again the the tribimaximal
and $m_\mu=(2/3)m_\tau$.

Case 3. The right-handed components of the charged leptons transform as in Case 1 but we introduce three Higgs doublets, $\Phi_e,\Phi_\mu,\Phi_\tau$,
 transforming under $S_3$ as a singlet $H_S=\Phi_e$ and a doublet $H_D=(\Phi_\mu,\Phi_\tau)$. The Yukawa interction is
\begin{eqnarray}
-\mathcal{L}_l&=&h_e\bar{L}_2e_SH_S+h_2[\bar{L}_2e_D]_2H_D]_1+h_3[[\bar{L}_DH_S]_2e_D]_1+h_4[[\bar{L}_DH_D]_2e_D]_1+h_5[[\bar{L}_DH_S]_2e_D]_1\nonumber \\&+&
h_6[\bar{L}_DH_D]_1e_S+H.c.
\label{ea3}
\end{eqnarray}
In this case we have that $\textrm{Det}M^l=0$, hence the electron remains massless at tree level.

Case 4. Introduce Higgs scalars as in Case 3 and all right-handed charged lepton transform as singlet under $S_3$. The Yukawa interactions is
\begin{equation}
-\mathcal{L}_l=(h_i\bar{L}_SH_S+h^\prime_i [\bar{L}_DH_D]_1)l_{iR}+H.c.
\label{ea4}
\end{equation}
In this case we obtain as in the Case 1, the trimaximal matrix in the charged lepton sector and two leptons remains massless at tree level.

Case 5. Now the $e_R,\mu_R,\tau_R$ transform as in Eq.~(\ref{ea2}) and the three Higgs doublets as in Case 3 and the Yukawa interactions are also given in Eq.~(\ref{ea3}).
This case is difficult to analyse analitically but numerical calculations indicate altought we can fit the three charged lepton masses the mixing matrix is given by
\begin{equation}
U^l_L=\left(
\begin{array}{ccc}
-0.382&0.621&0.684\\
0.360&0.582&0.730\\
0.851&0.525&0.001
\end{array}\right)
\label{ea5}
\end{equation}
and the PMNS matrix
\begin{equation}
\vert V_{PMNS}\vert=\left(\begin{array}{ccc}
0.51&0.06&0.86\\
0.06&0.99&0.04\\
0.86&0.03&0.52
\end{array}\right)
\label{ea6}
\end{equation}
does not fit the experimental values, see Eq.~(\ref{magic3}).

All these cases allow a diagonal mass matrix in the charged lepton sector if extra $Z_n$ symmetries are added as in Ref.~\cite{moha10}.


\end{document}